\documentclass[a4paper,doublecolumn]{IEEEtran}

\usepackage{epsfig,epstopdf,graphicx,subfigure,psfrag,amsmath,cases}
\usepackage{latexsym,amssymb,amsmath,epsfig,subfigure,algorithm,amsthm}
\usepackage{nopageno}
\usepackage{algorithmic}
\usepackage{color}
\usepackage{url}
\usepackage{scrtime}
\usepackage{bm}
\usepackage{graphicx}
\usepackage{bbding}
\usepackage{multicol}
\usepackage{graphicx}
\usepackage{amsthm}
\usepackage{cite}
\usepackage{array}
\usepackage[utf8]{inputenc}
\usepackage[english]{babel}
 \usepackage{setspace}
\addto\captionsenglish{}

\title{Sum-Rate Maximization for Multiuser MISO Downlink Systems with Self-sustainable IRS}

\author{\IEEEauthorblockN{Shaokang Hu, Zhiqiang Wei, Yuanxin Cai, Derrick Wing Kwan Ng, and Jinhong Yuan}\\
        \IEEEauthorblockA{School of Electrical Engineering \& Telecommunications, University of New South Wales, Sydney, Australia}\vspace{-3mm}}

{}
\newtheorem{theorem}{Theorem}

\begin{document}
\maketitle

\thispagestyle{empty}

\begin{abstract}
This paper investigates multiuser multi-input single-output (MISO) downlink communications assisted by a self-sustainable intelligent reflection surface (IRS), which can harvest power from the received signals. We study the joint design of the beamformer at an access point (AP) and the phase shifts and the power harvesting schedule at an IRS for maximizing the system sum-rate. The design is formulated as a non-convex optimization problem taking into account the capability of IRS elements to harvest wireless power for realizing self-sustainability. Subsequently, we propose a computationally-efficient alternating algorithm to obtain a suboptimal solution to the design problem. Our simulation results unveil that: 1) there is a non-trivial trade-off between the system sum-rate and self-sustainability in IRS-assisted systems; 2) the performance gain achieved by the proposed scheme is improved with an increasing number of IRS elements; 3) an IRS equipped with small bit-resolution discrete phase shifters is sufficient to achieve a considerable system sum-rate of an ideal case with continuous phase shifts.
\end{abstract}
\large\normalsize
\section{Introduction}
The sixth-generation (6G) networks are expected to serve as a key enabler for the future intelligent digital society in 2030, offering superior communication services compared with the current fifth-generation (5G) networks. It is foreseen that 6G networks will reach up to a connectivity density with $10^7$ devices/$\mathrm{km}^2$\cite{zhang20196g}. However, battery-powered wireless communication devices are equipped with limited energy storage that shortens the lifetime of communication networks. As a result, powering wirelessly connected devices to offer uninterrupted communication services will be an essential design challenge in 6G. In practice, wireless power transfer (WPT) is an effective solution\cite{clerckx2018fundamentals} to avoid manually replacing batteries of wireless devices, which may be costly or even impossible due to environmental hazards. In particular, harvesting power from radio frequency (RF) in wireless communication systems is more reliable than that from natural sources, e.g. wind, geothermal, and solar, due to the controllability of WPT.

To fulfill the stringent requirements set by 6G, such as ultra-low power consumption and high spectral efficiency, the emerging intelligent reflecting surface (IRS)-assisted wireless communications\cite{wu2019intelligent} have received considerable attentions recently. Specifically, an IRS consists of a large number of low-cost passive reflection elements that can independently reflect the incident electromagnetic wave with a particular phase shift. By intelligently adapting the phase shifts of each element at an IRS to the communication channels, the reflected signals can be coherently combined at the desired receivers. As such, IRS can establish a favorable communication environment for harnessing multiple access interference and enhancing the efficiency of communication. For example, \cite{wu2019intelligent} illustrated that IRS-assisted communication systems can extend the signal coverage compared with direct transmission in  conventional systems. Furthermore, \cite{wu2019weighted} showed that the introduction of an IRS can significantly improve both the achievable system data rate and the total harvested power in simultaneous wireless information and power transfer (SWIPT) systems. Besides, considering the impact of finite-resolution phase shifters at an IRS,  a joint design of beamforming and phase shifts was proposed to minimize the transmit power at an access point (AP)\cite{wu2019beamforming}. Despite the fruitful results in the literature, e.g. \cite{wu2019intelligent,wu2019weighted,wu2019beamforming}, most of the works idealistically assumed that the power consumption of the IRS is negligible as the IRS only contains passive elements. However, the power consumption of the IRS in practical systems is considerable compared to the transmit power \cite{huang2019reconfigurable}. To facilitate the design of energy-efficient IRS systems, a practical power consumption model of the IRS was proposed in \cite{huang2019reconfigurable}. In fact, the primary power consumption of an IRS arises from the feeding circuits to diodes for the reflection elements that depends on the bit resolution of the individual phase shifter of each reflection element. More importantly, the total power consumption of the IRS is proportional to the number of IRS elements and a massive number of reflecting elements are usually deployed to improve system performance. As a result, a self-sustainable IRS powered by WPT was considered in \cite{lyu2020optimized} for improving the system sum-rate of a hybrid-relaying scheme. However, \cite{lyu2020optimized}  assumed the availability of continuous phase shifters which is over optimistic for practical implementations due to the related hardware limitation and the associated cost. Moreover,  \cite{lyu2020optimized} focused on the resource allocation design of a single-antenna AP and the result cannot be applied to the case of multi-antenna transmitters. In fact, an efficient beamforming design to strike a balance between the system sum-rate and IRS self-sustainability has not been reported in the literature yet.

Motivated by the aforementioned observations, we consider a self-sustainable IRS-assisted multiuser MISO downlink wireless system, where the IRS is equipped with discrete phase shifters. In particular, our design advocates some of the IRS elements to harvest the received power for supporting the power consumption of the IRS such that the IRS does not require any extra power source. The precoding at the AP and the discrete phase shifts and power harvesting schedule at the IRS are jointly optimized to maximize the system sum-rate. The resource allocation design is formulated as a non-convex mixed-integer optimization problem, which is generally intractable. To tackle the design problem, we transform the sum-rate maximization problem into its equivalent form which facilitates the development of a computationally-efficient alternating optimization-based algorithm to obtain a suboptimal solution of the design problem. Our results not only show the non-trivial trade-off between the system sum-rate and self-sustainability of IRS-assisted systems, but also unveil the impact of bit resolution of the IRS phase shifters on the system performance.

\emph{Notations}: The scalars, vectors, and matrices are represented by lowercase letter $x$, boldface lowercase letter $\mathbf{x}$, and boldface uppercase letter $\mathbf{X}$, respectively. $\mathbb{R}^{N \times M}$ and $\mathbb{C}^{N \times M}$ denote the space of $N \times M$ matrices with real and complex entries, respectively.  $\mathbb{H}^N$ denotes the set of all $N\times N$ Hermitian matrices. The modulus of a complex-valued scalar and an Euclidean norm of a vector are denoted by $|\cdot|$ and $\|\cdot\|$, respectively. The transpose, conjugate transpose, conjugate, expectation, rank, and trace of a matrix are denoted as $(\cdot)^{\mathrm{T}}$, $(\cdot)^{\mathrm{H}}$, $(\cdot)^{*}$, $\mathbb{E}\{\cdot\}$, $\mathrm{Rank}(\cdot)$, and $\mathrm{Tr(\cdot)}$, respectively. $\mathbf{X}\succeq\mathbf{0}$ means that matrix $\mathbf{X}$ is positive semi-definite. $\mathrm{Diag(\mathbf{x})}$ denotes a diagonal matrix with its diagonal elements given by vector $\mathbf{x}\in \mathbb{C}^{N\times1}$. $j$ denotes the imaginary unit. For a continuous function $f(\mathbf{X})$, $\nabla_{\mathbf{X}}f(\cdot)$ represents the gradient of $f(\cdot)$ with respect to matrix $\mathbf{X}$. The distribution of a circularly symmetric complex Gaussian (CSCG) random variable with mean $\mu$ and variance $\sigma^2$ is denoted by $\mathcal{CN}(\mu, \sigma^2)$ and $\sim$ stands for ``distributed as''. $\mathbf{I}_N$ denotes an $N\times N$ identity matrix. 
\vspace{-2mm}
\section{System Model}\vspace{-1mm}
\begin{figure}[t]
  \centering
  \includegraphics[width=3in]{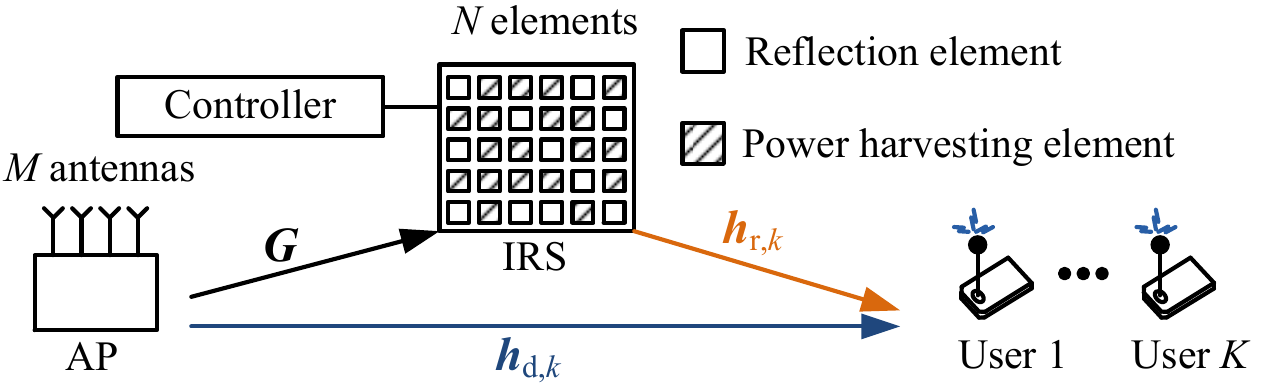}  \vspace*{-3mm}
  \caption{\hspace{-1mm}A downlink wireless communication system with a self-sustainable IRS.}
  \label{system_model} \vspace*{-5mm}
\end{figure}

As shown in Fig. \ref{system_model}, this paper considers a downlink MISO system assisted by a wireless-powered IRS. An AP equipping with $M>1$ antennas transmits $K$ independent data streams to $K$ single-antenna users simultaneously, denoted by a set $ \mathcal{K} = \{1, \ldots ,K\} $. The IRS panel consists of $N$ IRS elements, denoted by a set $\mathcal{N} = \{1, \ldots ,N\} $. The reflection matrix of the IRS is denoted as $\bm{\Theta}= \mathbf{A}\mathbf{\Phi}$, where $\mathbf{\Phi} = \mathrm{diag}(\beta_1 e^{j\theta_1},  \ldots,\beta_n  e^{j\theta_n} ,\ldots,\beta_N  e^{j\theta_N}) \in \mathbb{C}^{N\times N}$ is a diagonal matrix with phase shift $\theta_n \in [0,2\pi)$ and amplitude coefficient $\beta_n \in [0,1], \forall n \in \mathcal{N}$. Matrix $\mathbf{A}= \mathrm{diag}(\alpha_{1},\ldots,\alpha_{n},\ldots,\alpha_{N})\in \mathbb{R}^{N\times N}$, $\forall n \in \mathcal{N}$, and $\alpha_{n}\in\{0,1\}$ is an IRS mode selection variable which is defined as:
\begin{align}
    \alpha_{n} = \left\{
    \begin{array}{ll}
      \hspace{-2mm}  1, &\hspace{-2mm} \text{Reflection mode at IRS element} \,n\text{,}\\
      \hspace{-2mm}  0, &\hspace{-2mm} \text{Power harvesting mode at IRS element} \,n\text{.}
    \end{array}
\right.
\end{align}
For practical implementation of the IRS, the reflection coefficient $\beta_n $  is fixed to be $1$ in this paper, as commonly adopted in the literature, e.g.  \cite{wu2019intelligent,wu2019weighted,wu2019beamforming,huang2019reconfigurable}. On the other hand, each of the IRS elements can be scheduled operating in either reflection mode or power harvesting mode. Besides, we assume that discrete phase shifts are adopted in each IRS element and the phase shift interval $[0,2\pi)$ is uniformly quantized, i.e.,\vspace{-2mm}
\begin{align}
\theta_n \in \mathcal{F} = \mathcal{\mathfrak{}} \Big\{0, \ldots,\bigtriangleup\theta,\ldots, \bigtriangleup\theta(B-1)\Big\}, \forall n \in \mathcal{N},
\end{align}
where $\mathcal{F}$ is a set of phase shift, $\bigtriangleup\theta = 2\pi/B$, $B=2^b$ is the number of realizable phase shift levels, and $b$ is the given constant bit resolution. The amount of power consumed by each $b$-bit resolution reflection element is denoted by $P_{\mathrm{IRS}}(b)$\footnote{As the power consumption of an IRS is mainly dominated by the phase shifters, we assume that other energy consumptions, such as feedback or signaling overhead required by the IRS, are covered by $P_{\mathrm{IRS}}(b)$\cite{huang2019reconfigurable}.}. In particular, IRS elements in reflection mode reflect all impinging signal waveforms, while the elements in power harvesting mode harvest all the received power carried by the signals. Note that once an IRS element is in reflection mode, all the received signals are reflected and the IRS element cannot harvest any power at all. Likewise, the elements operating in power harvesting mode do not reflect any received signal.

This paper assumes a quasi-static flat fading channel model and the channel state information (CSI) for all links are assumed to be perfectly known at the AP. This can be achieved by applying existing CSI estimation algorithms\cite{wang2019channel}. The baseband equivalent channels from the AP to the IRS, from the IRS to the $k$-th user, and from the AP to the $k$-th user are denoted by $\bm{\mathbf{G}}\in\mathbb{C}^{N \times M}$, $\mathbf{h}_{\mathrm{r},k}\in\mathbb{C}^{N \times 1}$, and $\mathbf{h}_{\mathrm{d},k}\in\mathbb{C}^{M \times 1}$, respectively.
The transmitted signal from the AP is given by\vspace{-1mm}
\begin{align}
    \mathbf{x} = \sum_{k\in\mathcal{K}}\mathbf{w}_k x_k,
    \label{tx_signal}
\end{align}
where $\mathbf{w}_k\in\mathbb{C}^{M\times1}$ is the precoding vector for the $k$-th user and $x_k\sim \mathcal{CN}(0,1)$, $\forall k \in \mathcal{K}$, with $\mathbb{E}\{|x_k|^2\} = 1$, is the data symbol intended to the $k$-th user. We assume that the AP has a total transmit power $P_{\max}$, i.e., $\mathbb{E}\{\mathbf{x}\} = \sum_{k\in\mathcal{K}}\|\mathbf{w}_k\|^2\le P_{\max}$.
In the system, each user receives signals via two links, i.e., AP-user link and AP-IRS-user link. Thus, the signal received at the $k$-th user is given by\footnote{For a small-cell network with $200$ meters of cell radius, the delay between the propagation path reflected by the IRS and the direct path is typically around $1$ $\mathrm{\mu s}$, which is much shorter than a symbol duration, e.g. $70$ $\mathrm{\mu s}$ in Long-Term Evolution (LTE) systems\cite{arunabha2010fundamentals}. Therefore, the potential intersymbol interference caused by the two paths is not considered in \eqref{y}.}
\begin{align}
    y_k = \left(\mathbf{h}_{\mathrm{d},k}^\mathrm{H}+\mathbf{h}_{\mathrm{r},k}^\mathrm{H}\mathbf{A}\mathbf{\Phi}\mathbf{G}\right)\sum_{k\in \mathcal{K}}\mathbf{w}_k x_k +n_k,\label{y}
\end{align}
where $n_k \sim \mathcal{CN}(0,\sigma_k^2)$ is the background noise at the $k$-th user with a noise power $\sigma_k^2$. Accordingly, the received SINR at the $k$-th user, $\forall k \in \mathcal{K}$, is given by\vspace{-1mm}
\begin{align}
    \mathrm{SINR}_k
     =\frac{|(\mathbf{h}_{\mathrm{d},k}^\mathrm{H}+\mathbf{h}_{\mathrm{r},k}^\mathrm{H}\mathbf{A}\mathbf{\Phi}\mathbf{G})\mathbf{w}_k|^2}{\sigma_k^2+\sum_{j\neq k}|(\mathbf{h}_{\mathrm{d},k}^\mathrm{H}+\mathbf{h}_{\mathrm{r},k}^\mathrm{H}\mathbf{A}\mathbf{\Phi}\mathbf{G})\mathbf{w}_j|^2}\text{.}
\end{align}
The achievable rate (bits/s/Hz) of the $k$-th user is given by
\begin{align}
    R_k = \log_2\left(1+\mathrm{SINR}_k\right),\forall k \in \mathcal{K}\text{.}
\end{align}
Additionally, the total received signals for power harvesting at the IRS is given by\vspace*{-1mm}
\begin{align}
    \mathbf{y}_{\mathrm{EH}} (\mathbf{A},\mathbf{w}_k)=\mathbf{A}_{\mathrm{EH}}(\mathbf{G}\mathbf{x}+\mathbf{n}_a), \label{y_eh}
\end{align}
where $ \mathbf{A}_{\mathrm{EH}}=\mathbf{I}_N-\mathbf{A}$ is the power harvesting binary-valued matrix, $\mathbf{n}_a \in \mathbb{C}^{N \times 1}$, and $\mathbf{n}_a \sim \mathcal{CN}(\mathbf{0},\sigma_a^2\mathbf{I}_N)$ is the receiving thermal noise at the IRS with noise power per IRS element $\sigma_a^2$. The total harvested power by the IRS is given by\vspace*{-1mm}
\begin{align}
P_{\mathrm{EH}}=\eta_h\mathbb{E}\Big(\|\mathbf{A}_{\mathrm{EH}}(\mathbf{G}\sum_{k\in\mathcal{K}}\mathbf{w}_kx_k+\mathbf{n}_a)\|^2\Big),
\end{align}
where $0\leq\eta_h\leq1$ is the power harvesting efficiency of the IRS elements in converting the received RF signal into electrical energy.
\vspace{-3mm}
\section{Problem Formulation}\vspace*{-1mm}
We aim to maximize the system sum-rate while maintaining the self-sustainability of the IRS by jointly designing the precoding vector $\{\mathbf{w}_k\}_{k\in\mathcal{K}}$ at the AP, the mode selection $\{\alpha_n\}_{n\in\mathcal{N}}$, and the discrete phase shifter $\{\theta_n\}_{n\in\mathcal{N}} $ adopted at IRS. The joint design can be formulated as the following optimization problem\footnote{We note that the considered problem can be easily extended to the case of power harvesting users, at the expense of more involved notations.}:
\begin{align}
&\underset{\mathbf{w}_k,\, \alpha_n,\,\theta_n}{\mathrm{maximize}} \,\,\sum_{k\in\mathcal{K}} \log_2(1+\mathrm{SINR}_k)  \label{proposed_formulation_origion} \\
&\mathrm{s.t.}\,\,\mathrm{C1}\hspace{-1mm}: \sum_{k\in\mathcal{K}}\|\mathbf{w}_k\|^2 \leq P_{\max},\quad \mathrm{C2}\hspace{-1mm}: \theta_n \in \mathcal{F}, \forall n \in \mathcal{N}, \notag\\
&\hspace{5.5mm}\mathrm{C3}\hspace{-1mm}:  \sum_{n=1}^N \alpha_{n}P_{\mathrm{IRS}}(b) \leq P_{\mathrm{EH}}, \mathrm{C4}\hspace{-1mm}: \alpha_{n}\in \{0,1\},\forall n, \notag
\end{align}
where constraint C1 ensures that the transmit power at the AP does not exceed its maximum transmit power budget $P_{\max}$. Constraint C2 specifies that the phase shift of a $b$-bit resolution IRS reflecting element can only be selected from a discrete set $\mathcal{F}$. Constraint C3 indicates that the total power consumed at the IRS should not exceed its total harvested power from the AP, $P_{\mathrm{EH}}$. Constraint C4 is imposed to guarantee that each IRS element can only operate in either reflection mode or power harvesting mode. The formulated problem is non-convex due to the coupling between variables $\mathbf{w}_k$, $\theta_n$, and $\alpha_n$ in the sum rate expression, the discrete phase shift constraint C2, and the binary variable $\alpha_n$ constraint C3.  In general, finding the globally optimal solution of \eqref{proposed_formulation_origion} requires the application of a brute-force search which is computationally prohibited even for a moderate system size. As a compromise approach, in the sequel, we propose a computationally efficient suboptimal iterative algorithm based on alternating optimization.
\vspace{-2mm}
\section{Solution Of The Optimization Problem} \label{solution}

\subsection{Problem Transformation}
To facilitate the design of discrete IRS phase shifts, we first handle the coupling of $\mathbf{A}\mathbf{\Phi}$ in the objective function. To this end, we define an augmented mode selection matrix $\tilde{\mathbf{A}} = \mathrm{diag}(\tilde{\bm{\alpha}})=\mathrm{diag}(\tilde{\alpha}_1,$$ \ldots, \tilde{\alpha}_n, \ldots, $$\tilde{\alpha}_N)$ with $B+1$ modes, where $\tilde{\alpha}_n\in \tilde{\mathcal{F}} =\{0,e^{j0},$$\ldots,e^{j\bigtriangleup\theta},\ldots, $$e^{j\bigtriangleup\theta(B-1)}\}$ is the mode selection of the $n$-th element and $\tilde{\mathcal{F}}$ is the generalized mode selection set. When $\tilde{\alpha}_{n} = 0$, the $n$-th IRS element is in the power harvesting mode, otherwise it is in the reflection mode. Therefore, constraints C3 and C4 in \eqref{proposed_formulation_origion} can be equivalently rewritten as:\vspace*{-1mm}
\begin{align}
&\mathrm{C3}\hspace{-1mm}:\sum_{n=1}^N |\tilde{\alpha}_{n}|P_{\mathrm{IRS}}(b) \leq P_{\mathrm{EH}} \text{ and} \\
&\mathrm{C4}\hspace{-1mm}:\tilde{\alpha}_{n} \in \tilde{\mathcal{F}} =\{0,e^{j0}, e^{j\bigtriangleup\theta},\ldots, e^{j\bigtriangleup\theta(B-1)}\},\forall n \text{,} \label{c4}
\end{align}
respectively. Then, to handle the discrete variable $\tilde{\alpha}_{n}$ in C3 and C4, we further introduce a binary mode selection optimization variable $s_{i,n}, \forall i \in \mathcal{I}= \{1,\ldots,B+1\}, n\in \mathcal{N}$, and a mode selection binary matrix $\mathbf{S}\in \mathbb{R}^{(B+1)\times N}$, $s_{i,n} \in \mathbf{S}$. In particular, $s_{i,n} = 1$ indicates that the $i$-th mode is selected for the $n$-th element. Otherwise, $s_{i,n}=0$. Thus, constraint C4 can be represented as:\vspace{-1mm}
\begin{align}
&\mathrm{C4a}\hspace{-1mm}:\sum_{i\in\mathcal{I}}s_{i,n} \le 1, \forall n, \hspace{3mm}\mathrm{C4b}\hspace{-1mm}:\tilde{\alpha}_{n} = \sum_{i\in\mathcal{I}}s_{i,n}f_i, \forall n,\label{ModeEH}\\
&\mathrm{C4c}\hspace{-1mm}:s_{i,n} \in \{0,1\}, \forall i,n,
\end{align}
where $f_i$ is the $i$-th element of the generalized mode selection set $\tilde{\mathcal{F}}$ defined in \eqref{c4}. Meanwhile, $\mathbf{A}_{\mathrm{EH}}$ in \eqref{y_eh} can be rewritten as a function of $\mathbf{s}_1$, i.e., mode $1$ in \eqref{ModeEH}, which is given by\vspace{-1mm}
\begin{align}
\mathbf{A}_{\mathrm{EH}}= \mathrm{diag}(\mathbf{s}_1),
\end{align}
where $\mathbf{s}_1 = [s_{1,1},\ldots,s_{1,n},\ldots,s_{1,N}]^{\mathrm{T}}$ is the transpose of the first row of mode selection binary matrix $\mathbf{S}$.
Thus, constraint C3 can be equivalently rewritten as\vspace{-2mm}
\begin{align}
   \hspace{-1mm}\mathrm{\overline{C3}}\hspace{-1mm}: \hspace{1mm}&(N- \sum_{n=1}^N (s_{1,n}))P_{\mathrm{IRS}}(b) \notag\\ &\hspace{-1mm}\leq\hspace{-1mm}\eta_h\Big(\hspace{-1mm}\sum_{k\in\mathcal{K}}\hspace{-1mm}\mathrm{Tr}(\mathbf{G}\mathbf{w}_k\mathbf{w}_k^\mathrm{H}\mathbf{G}^\mathrm{H} \mathrm{diag}(\mathbf{s}_1))+\sigma_a^2\hspace{-1mm}\sum_{n=1}^N s_{1,n}\Big).\label{barc3}
\end{align}
It can be seen from \eqref{barc3} that there is a non-trivial trade-off between the system sum-rate and the number of IRS elements in power harvesting mode. To achieve the self-sustainability of the IRS, some of the IRS elements are exploited for harvesting power leading to a smaller number of IRS elements for signal reflection to improve the system sum-rate.
Now, the problem in \eqref{proposed_formulation_origion} can be equivalently transformed to\vspace{-2mm}
\begin{align}
&\underset{\mathbf{w}_k,\mathbf{S},\tilde{\alpha}_{{n}}} {\mathrm{maximize}}   \,\, \sum_{k\in\mathcal{K}} \log_2(1+ \mathrm{SINR}_k) \label{optimization_problem}\\
&\mathrm{s.t.} \,\, \mathrm{C1},
\mathrm{\overline{C3}},
\mathrm{C4a},
\mathrm{C4b},
\mathrm{C4c}.\notag
\end{align}
In the following, we focus on solving the optimization problem in \eqref{optimization_problem}. Note that although the problem in \eqref{optimization_problem} is still non-convex, the above transformation facilitates the application of alternating optimization for achieving a suboptimal solution. In particular, the proposed algorithm tackles the coupling variables $\mathbf{w}_k$ and $\{\mathbf{S} ,\tilde{\alpha}_n,\forall n\}$ by dividing \eqref{optimization_problem} into two subproblems, i.e., we alternatingly update $\{\mathbf{S},\tilde{\alpha}_n,\forall n\}$ and $\{\mathbf{w}_k,\forall k\}$ while the other variables are fixed in solving the two subproblems, respectively.\vspace{-4mm}
\subsection{Sub-problem 1: Optimization of Precoder at the AP}\label{Optimization_of_Transmit_Beamforming}\label{subp1}\vspace{-1mm}
In this section, we aim to optimize the transmit beamforming vector $\mathbf{w}_k$, for a given fixed feasible point $\{\mathbf{S}^{\mathrm{con.}},\tilde{\alpha}_n^{\mathrm{con.}},\forall n\}$, where $\mathbf{S}^{\mathrm{con.}}$ is the mode selection binary matrix with fixed values and $\tilde{\alpha}_n^{\mathrm{con.}},\forall n$, is the mode selection for the $n$-th element with a fixed value. By defining $\mathbf{W}_k \triangleq \mathbf{w}_k\mathbf{w}_k^H$, the problem in \eqref{optimization_problem} can be rewritten as\vspace{-2mm}
\begin{align}
&\underset{\mathbf{W}_k\in\mathbb{H}^M}{\mathrm{minimize}}\, -\hspace{-2mm}\sum_{k\in\mathcal{K}} \log_2\Big(1+ \frac{\mathrm{Tr}(\mathbf{W}_k\mathbf{M}_k)}{\sigma_k^2+\sum_{j\neq k}\mathrm{Tr}(\mathbf{W}_j\mathbf{M}_k)}\Big)   \label{Afix_problem}\\
&\mathrm{s.t.} \,\,\mathrm{C1}, \mathrm{\overline{C3}}, \mathrm{C5}\hspace{-1mm}:\mathbf{W}_k \succeq \mathbf{0}, \forall k, \mathrm{C6}\hspace{-1mm}: \mathrm{Rank}(\mathbf{W}_k)\le 1,\forall k,\notag
\end{align}
where $\mathbf{M}_k = \mathbf{m}_k\mathbf{m}_k^\mathrm{H}, \mathbf{m}_k= \mathbf{h}_{\mathrm{d},k}+\mathbf{G}^\mathrm{H}\tilde{\mathbf{A}}^\mathrm{H}\mathbf{h}_{\mathrm{r},k}$. Constraints C5, C6, and $\mathbf{W}_k\in\mathbb{H}^M$ are imposed to guarantee that $\mathbf{W}_k=\mathbf{w}_k\mathbf{w}_k^\mathrm{H}$ still holds after optimizing $\mathbf{W}_k$. Now, we apply an iterative method based on the successive convex approximation (SCA) to tackle the non-convexity of the objective function in \eqref{Afix_problem}. To start with, we first rewrite the objective function of \eqref{Afix_problem} in the form of difference of convex (d.c.) functions\cite{yu2019robust}:\vspace{-1mm}
\begin{align}
    \hspace{-3mm}-\sum_{k\in\mathcal{K}} \log_2\Big(1+ \frac{\mathrm{Tr}(\mathbf{W}_k\mathbf{M}_k)}{\sigma_k^2+\sum_{j\neq k}\mathrm{Tr}(\mathbf{W}_j\mathbf{M}_k)}\Big) = N_1-D_1,
\end{align}
where
\vspace{-2mm}
\begin{align}
    N_1 & = -\sum_{k\in\mathcal{K}} \log_2(\sigma_k^2+\sum_{j\in\mathcal{K}}\mathrm{Tr}(\mathbf{W}_j\mathbf{M}_k))\,\, \text{and}\\
    D_1 & = -\sum_{k\in\mathcal{K}} \log_2(\sigma_k^2+\sum_{j\neq k}\mathrm{Tr}(\mathbf{W}_j\mathbf{M}_k))
\end{align}
are two functions that are both convex with respect to $\mathbf{W}_k$. For any feasible point $\mathbf{W}_k^{t^{(\mathrm{1})}},\forall k\in\mathcal{K}$, where $t^{(\mathrm{1})}$ denotes the iteration index for \textbf{Algorithm \ref{table:alg_1}}, a lower bound function of $D_1$ is given by its first-order Taylor expansion:\vspace{-1mm}
\begin{align}
  D_1(\mathbf{W}_k) \geq& \sum_{k\in\mathcal{K}}\mathrm{Tr}\Big(\nabla^\mathrm{H}_{\mathbf{W}_k}D_1(\mathbf{W}_k^{t^{(\mathrm{1})}})(\mathbf{W}_k-\mathbf{W}_k^{t^{(\mathrm{1})}})\Big)\notag\\
 & + D_1(\mathbf{W}_k^{t^{(\mathrm{1})}}),\label{D1}
\end{align}
where the first partial derivative of function $D_1$ with respect to $\mathbf{W}_k$ is given by\vspace{-1mm}
\begin{align}
\hspace{-2mm}\nabla_{\mathbf{W}_k}D_1(\mathbf{W}_k)\hspace{-1mm} = \hspace{-1mm}\frac{-1}{\ln2}\hspace{-1mm}\sum_{j\ne k}\hspace{-1mm}\Big(\frac{\mathbf{M}_j}{\sigma^2_k+\sum_{q\in \mathcal{K}\backslash\{j\}}\hspace{-1mm}\mathrm{Tr}(\mathbf{W}_q\mathbf{M}_j)}\Big).\hspace{-2mm}\hspace{-1mm}
\end{align}
By replacing $D_1$ in the objective function of \eqref{Afix_problem} with \eqref{D1}, an upper bound problem of \eqref{Afix_problem} is obtained.
Now, the rank constraint C6 is the only non-convexity of the problem. To circumvent this issue, we adopt the semidefinite relaxation (SDR) technique \cite{wu2019weighted} and drop the rank constraint. Therefore, the resulting optimization problem is given by\vspace{-1mm}
\begin{align}
\underset{\mathbf{W}_k\in\mathbb{H}^M}{\mathrm{minimize}}  & \,\,N_1-\sum_{k\in\mathcal{K}}\mathrm{Tr}\Big(\nabla^\mathrm{H}_{\mathbf{W}_k}D_1(\mathbf{W}^{t^{(\mathrm{1})}})(\mathbf{W}_k-\mathbf{W}_k^{t^{(\mathrm{1})}})\Big) \notag\\
&+\sum_{k\in\mathcal{K}} \log_2\Big(\sigma_k^2+\sum_{j\neq k}\mathrm{Tr}(\mathbf{W}_j^{t^{(\mathrm{1})}}\mathbf{M}_k)\Big)\label{subproblem1_final}\\
\mathrm{s.t.} &\,\, \mathrm{C1},
\mathrm{\overline{C3}},
\mathrm{C5}.\notag
\end{align}
Now, problem \eqref{subproblem1_final} is a convex semidefinite programming that can be solved by some standard convex program solvers. In the following theorem, we study the tightness of the adopted SDR.\vspace{-1mm}
\begin{theorem}
For $P_{\max}>0$ and if \eqref{subproblem1_final} is feasible, a rank-one solution of \eqref{subproblem1_final} can always be constructed.
\end{theorem}
\emph{\quad Proof: }
Due to page limitation, we only provide a sketch of the proof. By analyzing the Karush-Kuhn-Tucker (KKT) conditions of \eqref{subproblem1_final}, one can show that a rank-one solution $\mathbf{W}_k$ must exist to have a bounded dual problem solution of \eqref{subproblem1_final}. Besides, we can construct a rank-one solution of \eqref{subproblem1_final} by exploiting the dual variables of the dual problem of \eqref{subproblem1_final}.
 \qed

Due to the use of SCA, solving the problem in \eqref{subproblem1_final} provides an upper bound for the problem in \eqref{Afix_problem}. To tighten the obtained upper bound, we iteratively update the feasible solution $\mathbf{W}_k$ by solving the optimization problem in \eqref{subproblem1_final} in $t^{(\mathrm{1})}$ iteration. The proposed SCA-based algorithm is shown in \textbf{Algorithm \ref{table:alg_1}} and the proof of its convergence to a suboptimal solution can be found in \cite{opial1967weak} which is omitted here for brevity.\begin{table}[t]
\vspace*{-3.9mm}
\renewcommand\thealgorithm{1}
\begin{algorithm}[H]
\caption{SCA-based Iterative AP Precoder Design}\label{table:alg_1}        
\begin{algorithmic}[1]                     
\STATE Initialize the maximum number of iteration $t_{\max}^{(\mathrm{1})}$, the initial iteration index $t^{(\mathrm{1})}=0$, and variable $\{\mathbf{W}_k^{t^{(\mathrm{1})}}\}$ for given constants $\{s_{i,n}^{\mathrm{con.}},\forall i,n\} $, and $\{\tilde{\alpha}_n^{\mathrm{con.}}, \forall n\}$;
\REPEAT [Main loop]
\STATE Solve problem \eqref{subproblem1_final} with a given $\mathbf{W}_k^{t^{(\mathrm{1})}}$, $\{s_{i,n}^{\mathrm{con.}},\forall i,n\}$, and $\{\tilde{\alpha}_n^{\mathrm{con.}}, \forall n\}$, to obtain $\mathbf{W}_k^{t^{(\mathrm{1})}+1}$;
\STATE Set $t^{(\mathrm{1})}=t^{(\mathrm{1})}+1$;
\UNTIL convergence  or $t^{(\mathrm{1})}=t^{(\mathrm{1})}_{\max}$.\vspace*{-1mm}
\end{algorithmic}
\end{algorithm}
\normalsize\vspace*{-9.2mm}
\end{table}
\vspace{-2mm}
\subsection{Sub-problem 2: Optimization of IRS Mode Selection and Phase Shifts}
In this subproblem, we aim to optimize the mode selection matrix $\mathbf{S}\hspace{-1mm} =\hspace{-1mm} \{s_{i,n},\forall i,n\}$ while fixing the transmit precoder $\{\mathbf{w}_k^{\mathrm{con.}}, \forall k\}$. First, we tackle the binary variable $s_{i,n}$ by equivalently transforming constraint C4c to the following  two constraints:\vspace{-3mm}
\begin{align}
&\overline{\mathrm{C4c}}\hspace{-1mm}: s_{i,n}-s^2_{i,n}\le0,\forall  i,n\text{, and} \\
&\mathrm{C4d}\hspace{-1mm}:  0\le s_{i,n}\le1,\forall  i,n,
\end{align}
where $s_{i,n},\forall i,n$, are continuous variables. For the ease of presentation, let $\mathbf{L}_{k} = \mathrm{diag}(\mathbf{h}_{\mathrm{r},k}^\mathrm{H})\mathbf{G}$. Then by applying $\mathbf{h}_{\mathrm{r},k}^\mathrm{H} \tilde{\mathbf{A}}\mathbf{G}= \mathbf{v}^{\mathrm{H}} \mathbf{L}_{k}$, where $\mathbf{v} = [\tilde{\alpha}_1, \ldots,\tilde{\alpha}_n,\ldots\tilde{\alpha}_N]^\mathrm{H}$, we have $|(\mathbf{h}^{\mathrm{H}}_{\mathrm{d},k}+\mathbf{h}^{\mathrm{H}}_{\mathrm{r},k}\tilde{\mathbf{A}}\mathbf{G})\mathbf{w}_k|^2=|\mathbf{h}^{\mathrm{H}}_{\mathrm{d},k}\mathbf{w}_k+\mathbf{v}^{\mathrm{H}}\mathbf{L}_{k}\mathbf{w}_k|^2$.
Now, sub-problem 2 can be reformulated as\vspace{-2mm}
\begin{align}
\underset{\mathbf{S}, \mathbf{v}, \xi_{k},\iota_{k}}{\mathrm{minimize}}  \,\,& -\sum_{k\in\mathcal{K}} \log_2\Big(1+ \frac{\xi_{k}}{\sigma_k^2+\iota_{k}}\Big) \label{subproblem2_2}\\
\mathrm{s.t.} \,\, &\mathrm{\overline{C3}},
\mathrm{C4a},
\mathrm{C4b},
\overline{\mathrm{C4c}},
\mathrm{C4d},\notag\\
&\mathrm{C7}\hspace{-1mm}:  \xi_{k} \leq |\mathbf{h}^{\mathrm{H}}_{\mathrm{d},k}\mathbf{w}_k+\mathbf{v}^{\mathrm{H}}\mathbf{L}_{k}\mathbf{w}_k|^2, \forall k, \notag\\
&\mathrm{C8}\hspace{-1mm}: \iota_{k} \geq \sum_{j\neq k}|\mathbf{h}^{\mathrm{H}}_{\mathrm{d},k}\mathbf{w}_j+\mathbf{v}^{\mathrm{H}}\mathbf{L}_{k}\mathbf{w}_j|^2, \forall k\text{,}\notag
\end{align}
where $ \xi_{k}\hspace{-1mm}=\hspace{-1mm}|\mathbf{h}^{\mathrm{H}}_{\mathrm{d},k}\mathbf{w}_k+\mathbf{v}^{\mathrm{H}}\mathbf{L}_{k}\mathbf{w}_k|^2, \forall k$, and $\iota_{k} = \sum_{j\neq k}|\mathbf{h}^{\mathrm{H}}_{\mathrm{d},k}\mathbf{w}_j+\mathbf{v}^{\mathrm{H}}\mathbf{L}_{k}\mathbf{w}_j|^2,\forall k$, are slack optimization variables. Note that the inequality constraints $\mathrm{C7}$ and $\mathrm{C8}$ are always satisfied with equality at the optimal solution of \eqref{subproblem2_2}. It can be observed that the objective function, $\overline{\mathrm{C4c}}$, and $\mathrm{C7}$ are standard d.c. functions. By following the same approach as for handling sub-problem 1 in Section \ref{subp1}, we apply the SCA to address the non-convexity in the objective function, $\overline{\mathrm{C4c}}$, and $\overline{\mathrm{C7}}$. Defining $t^{(\mathrm{2})}$ as the iteration index for \textbf{Algorithm \ref{table:alg_2}}, an upper bound of \eqref{subproblem2_2} can be obtained via solving the  following optimization problem:\vspace{-2mm}
\begin{align}
&\hspace*{-2mm}\underset{\mathbf{S}, \mathbf{v},\xi_{k},\iota_{k}}{\mathrm{minimize}} \,N_2\hspace{-1mm}-\hspace{-1mm}\sum_{k\in\mathcal{K}}\hspace{-1mm}\nabla_{\iota_{k}}^{\mathrm{H}}D_2(\iota_{k}^{t^{(\mathrm{2})\hspace*{-0.5mm}}})\Big(\iota_{k}-\iota_{k}^{t^{(\mathrm{2})}}\Big)\hspace*{-1mm}-\hspace*{-1mm}D_2(\iota_{k}^{t^{(\mathrm{2})}})\hspace{-2mm}\label{subproblem2_final}\\
&\mathrm{s.t.} \, \mathrm{\overline{C3}},
%
\mathrm{C4a},
\mathrm{C4b},
\overline{\overline{\mathrm{C4c}}},
\mathrm{C4d},
\mathrm{\overline{C7}},
\mathrm{C8},\notag
\end{align}
where \vspace{-2mm}\begin{align}
&\hspace{-3mm}N_2\hspace{-1mm} =\hspace{-1mm}-\hspace{-2mm}\sum_{k\in\mathcal{K}} \log_2\hspace{-1mm}\big(\sigma_k^2+\xi_{k}+\iota_{k}\big),\hspace{1mm}\nabla_{\iota_{k}} D_2(\iota_{k})\hspace{-1mm} =\hspace{-1mm} \frac{-1}{(\ln2)(\sigma_k^2+\iota_k)}\hspace{-1mm}\hspace{-1mm}\\
&\hspace{-3mm}D_2(\iota_{k}^{t^{(\mathrm{2})}})\hspace{-1mm} = -\sum_{k\in\mathcal{K}} \log_2\big(\sigma_k^2+\iota_{k}^{t^{(\mathrm{2})}}\big)\text{,}\\
&\hspace{-3mm}\overline{\overline{\mathrm{C4c}}}\hspace{-1mm}:\,s_{i,n}\hspace{-1mm}-\hspace{-1mm}(s_{i,n}^{t^{(\mathrm{2})}})^2\hspace{-1mm}-2s_{i,n}^{t^{(\mathrm{2})}}(s_{i,n}-s^{t^{(\mathrm{2})}}_{i,n})\hspace{-1mm}\le\hspace{-1mm}0,\forall  i,n \text{, and}\\
&\hspace{-3mm}\mathrm{\overline{C7}}\hspace{-1mm}: \,\xi_{k}\hspace{-1mm}-(\mathbf{h}^{\mathrm{H}}_{\mathrm{d},k}\mathbf{W}_k\mathbf{h}_{\mathrm{d},k}
+\mathbf{h}^{\mathrm{H}}_{\mathrm{d},k}\mathbf{W}_k\mathbf{L}_k^\mathrm{H}\mathbf{v}
+\mathbf{v}^{\mathrm{H}}\mathbf{L}_{k}\mathbf{W}_k\mathbf{h}_{\mathrm{d},k})\notag \\
&\,\,\,\,\,\,\,\,\,\,\,\,-\Big((\mathbf{v}^{t^{\mathrm{(\mathrm{2})}}})^{\mathrm{H}}\mathbf{L}_{k}\mathbf{W}_k\mathbf{L}_k^\mathrm{H}\mathbf{v}^{t^{\mathrm{(\mathrm{2})}}} \notag \\
&\,\,\,\,\,\,\,\,\,\,\,\,+2(\mathbf{v}^{t^{(\mathrm{2})}})^{\mathrm{H}}\mathbf{L}_k\mathbf{W}_k^\mathrm{H}\mathbf{L}_k^\mathrm{H}
    (\mathbf{v}-\mathbf{v}^{t^{(\mathrm{2})}})\Big)\leq 0,\forall k.
\end{align}
\begin{table}[t]
\vspace*{-3.9mm}
\renewcommand\thealgorithm{2}
\begin{algorithm}[H]

\caption{IRS Mode Selection and Phase Control}          
\label{table:alg_2}
\begin{algorithmic}[1]                     
\STATE Initialize the maximum number of iteration $t^{(\mathrm{2})}_{\max}$ and the initial iteration index $t^{(\mathrm{2})}=0$.
\STATE Given $\{\mathbf{W}_k^{\mathrm{\mathrm{con.}}}, \forall k\}$. Initialize variables $\{s_{i,n}^{t^{(\mathrm{2})}},\forall i,n\}$, $\{v_n^{t^{(\mathrm{2})}} = \sum_{i\in\tilde{\mathcal{F}}}s_{i,n}^{t^{(\mathrm{2})}}f_i^*, \forall n\}$, $\{\xi_{k}^{t^{(\mathrm{2})}}, \forall k\}$, and $\{\iota_{k}^{t^{(\mathrm{2})}}, \forall k,j\}$;
\REPEAT [Main loop]
\STATE Obtain $\{s_{i,n}^{t^{(\mathrm{2})}+1}, \forall i,n\}$, $\{v_n^{t^{(\mathrm{2})}+1}, \forall n\}$, $\{\xi_{k}^{t^{(\mathrm{2})}+1}, \forall k\}$, and $\{\iota_{k}^{t^{(\mathrm{2})}+1},\forall k\}$ with given $\{\mathbf{W}_k^{\mathrm{\mathrm{con.}}},\forall k\}$, $ \{\iota_{k}^{t^{(\mathrm{2})}}, \forall k\}$, $\{\xi_{k}^{t^{(\mathrm{2})}+1}, \forall k\}$ and $\{v_n^{t^{(\mathrm{2})}}\}$ by solving problem \eqref{subproblem2_final};
\STATE Set $t^{(\mathrm{2})}=t^{(\mathrm{2})}+1$;
\UNTIL convergence  or $t^{(\mathrm{2})}=t^{(\mathrm{2})}_{\max}$.
\end{algorithmic}
\end{algorithm}
\normalsize\vspace*{-9.2mm}
\end{table}\begin{table}[t]
\renewcommand\thealgorithm{3}
\begin{algorithm}[H]
\caption{Alternating Optimization Algorithm}\label{alg_overall}         
\begin{algorithmic}[1]                     
\STATE Initialize the maximum number of iteration $t^{(\mathrm{3})}_{\max}$, the initial iteration index $t^{(\mathrm{3})}=0$, variables $\{\mathbf{w}_k^{t^{(\mathrm{3})}}, \forall k\}$ and $\{s_{i,n}^{t^{(\mathrm{3})}}, \forall i,n\}$.
\REPEAT [Main loop]
\STATE Obtain $\mathbf{W}_k^{t^{(\mathrm{3})}+1}$ by \textbf{Algorithm \ref{table:alg_1}} with given $\mathbf{W}_k^{t^{(\mathrm{3})}}$, $\{s_{i,n}^{t^{(\mathrm{3})}},\forall i,n\}$, and $\{\tilde{{\alpha}}_n^{t^{(\mathrm{3})}+1},\forall n\}$;
\STATE Obtain $\{s_{i,n}^{t^{(\mathrm{3})}+1}, \forall i,n\}$ and $\{v_n^{t^{(\mathrm{3})}+1}, \forall n\}$ by \textbf{Algorithm \ref{table:alg_2}} with given $\{\mathbf{W}_k^{t^{(\mathrm{3})}+1}\}$ and $\{v_n^{t^{(\mathrm{3})}}\}$;
\STATE Update $\{\tilde{{\alpha}}_n^{t^{(\mathrm{3})}+1},\forall n\}$ and $\{v_n^{t^{(\mathrm{3})}+1},\forall n\}$; 
\STATE Set $t^{(\mathrm{3})}=t^{(\mathrm{3})}+1$;
\UNTIL convergence  or $t^{(\mathrm{3})}=t^{(\mathrm{3})}_{\max}$.
\end{algorithmic}
\end{algorithm}
\normalsize\vspace*{-9mm}
\end{table}Constraints $\overline{\overline{\mathrm{C4c}}}$ and $\mathrm{\overline{C7}}$ are a subset of $\overline{\mathrm{C4c}}$ and $\mathrm{C7}$, respectively due to the application of SCA. Then, the obtained upper bound of the problem in \eqref{subproblem2_2} is tightened by iteratively updating the feasible solutions $\{\mathbf{S},\mathbf{v},\xi_{k},\iota_{k}\}$ via solving the problem in \eqref{subproblem2_final} with a convex programming solver. The proposed algorithm for handling \eqref{subproblem2_final} is shown in \textbf{Algorithm \ref{table:alg_2}} and the overall algorithm is summarized in \textbf{Algorithm \ref{alg_overall}} which solves the two subproblems in \eqref{subproblem1_final}  and \eqref{subproblem2_final} iteratively. Note that the convergence of \textbf{Algorithm \ref{alg_overall}} to a suboptimal solution of \eqref{optimization_problem} is guaranteed with a polynomial time computational complexity\cite{yu2019robust}.
\vspace{-4mm}
\section{Numerical Results}

\begin{figure}[t] \vspace*{-1mm}
  \centering
  \includegraphics[width=3.0in]{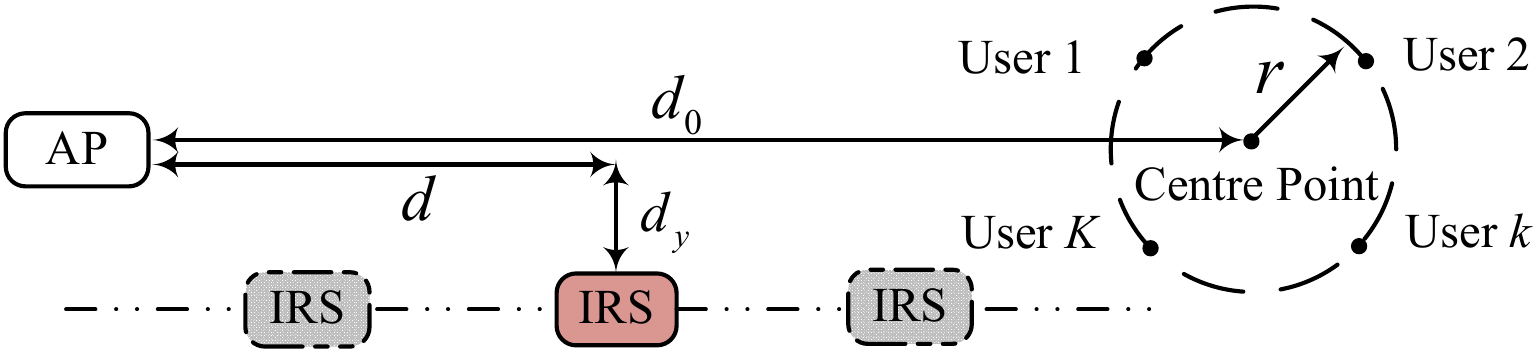}\vspace*{-2mm}
  \caption{Simulation setup.}
  \label{distance_model} \vspace*{-5mm}
\end{figure}

This section evaluates the system performance of the proposed self-sustainable IRS scheme via simulation. The system setup is shown in Fig. \ref{distance_model}. The users are randomly distributed on the circumference of a circle centered at a center point with radius $r = 1$ m. The AP and the center point of the circle are located $d_0 = 60$ m apart. The IRS is located between the AP and the center point with a vertical distance $d_y = 1$ m and a horizontal distance $d$ from the AP. The AP is equipped with a uniform linear array with $M=8$ antennas. The IRS is constituted by a uniform rectangular array with $N=256$ elements and there are $K=2$ users. The distance-dependent path loss model\cite{goldsmith2005wireless} is adopted with $10$ dBi transmit and receive antenna gains at the AP and the IRS, respectively, and $0$ dBi antenna gain at each user\cite{dai2020reconfigurable}.
The reference distance of the path loss model is $10$ meters. The system bandwidth is $200$ kHz and the carrier center frequency is $470$ MHz\cite{chen2010mac}. Due to the relatively long distance and random scattering of the AP-user channel, we set the path loss exponents of AP-user link as $\alpha_{\mathrm{AU}}=3.6$. Since the IRS is usually deployed to establish a line-of-sight (LoS) channel with the AP, we set the path loss exponents of AP-IRS link and IRS-user link as $\alpha_{\mathrm{AI}}=\alpha_{\mathrm{IU}}=2.2$. The small scale fading coefficients of the AP-user link, the AP-IRS link, and the IRS-user link are generated as independent and identically distributed (i.i.d.) Rican random variables with Rician factors $\beta_{\mathrm{AU}} = 0$, $\beta_{\mathrm{AI}} = 2$, and $\beta_{\mathrm{IU}} = 2$, respectively. We assume that the signal processing noise in each receiver is caused by thermal noise and quantization noise. Specifically, a $12$-bit uniform quantizer quantizes the received information at the receiver each user. As a result, for each user, the thermal noise and the quantization noise powers are -$110$ dBm and -$47$ dBm \cite{393000}, respectively.
Other important parameters are summarized as follows unless specified otherwise: the maximum power budget at the AP is $P_{\max} = 38$ dBm, the phase shifter bit resolution of each IRS reflection element $b = 3$ bits with power consumption  $P_{\mathrm{IRS}}(b) = 1$ dBm \cite{huang2019reconfigurable}, and the power harvesting efficiency of IRS elements $\eta_h=0.8$.

For comparison, we also evaluate the system performance of three other schemes: 1) A performance upper bound achieved by an IRS-assisted system with an idealistic IRS, e.g. all IRS elements in this scheme are in the reflection mode but without consuming any power; 2) Baseline scheme $1$ is designed for the case when IRS is not deployed. In particular, maximum ratio transmission (MRT) with respect to users is adopted for the precoder at the AP. The direction of the precoder for the $k$-th user is fixed to $\frac{\mathbf{h}_{\mathrm{d},k}^{\mathrm{H}}}{\|\mathbf{h}_{\mathrm{d},k}\|}$ and we optimize the power of the precoder of each user subject to constraint C1 in problem \eqref{proposed_formulation_origion} for the maximization of system sum-rate; 3) Baseline scheme $2$ is the system with a self-sustainable IRS adopting the same MRT precoder at the AP as in baseline scheme $1$. Its phase shifts of the IRS and the precoder power allocation at the AP are jointly optimized by \textbf{Algorithm \ref{alg_overall}}.

\begin{figure}[t] \vspace*{-5mm}
  \centering
  \includegraphics[width=3.5in]{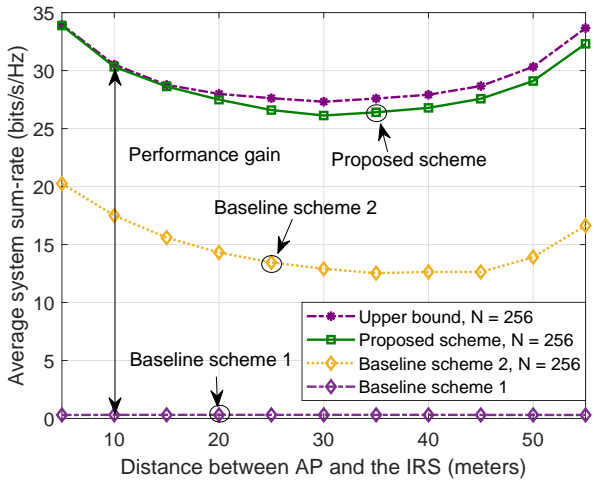} \vspace*{-3mm}
  \caption{Average system sum-rate (bits/s/Hz) versus the horizontal distance of AP-IRS link.}
  \label{results:distanceSR}\vspace*{-4mm}
\end{figure}

Fig. \ref{results:distanceSR} depicts the average system sum-rate versus the horizontal distance for different schemes. It can be observed that both the proposed scheme and the upper bound scheme achieve a substantially higher sum-rate than that of baseline scheme $1$. Indeed, the IRS provides an additional path gain, $\mathbf{h}_{\mathrm{r},k}^{\mathrm{H}}\mathbf{\Theta}\mathbf{G},\forall k$, which carries the same useful information as the direct link to the users. More importantly, this additional path gain is exploited and optimized by the proposed scheme to improve the system performance. Besides, due to the joint optimization of the precoder and phase shifts, the proposed scheme can achieve a considerable performance gain compared with baseline scheme $2$. On the other hand, for all schemes with IRS deployed, the average system sum-rate is at its lowest when the IRS is close to the middle between the AP and the center point of users. In fact, when the IRS is neither close to the AP nor the users, both the AP-IRS path and the IRS-user paths would experience significant attenuations that decreases the capability of the IRS in focusing the reflected signals on the desired users. We can also observe from Fig. \ref{results:distanceSR} that when the IRS is in close proximity to the AP, the performance of the proposed scheme approaches that of the upper bound. In contrast, as the distance $d$ further increases, the sum-rate gap between the proposed scheme and the upper bound is slightly larger. This is because as $d$ increases, each IRS element would harvest less power on average. As can be expected from constraint $\overline{\mathrm{C}3}$ in \eqref{barc3}, more IRS elements are switched to the power harvesting mode to maintain the sustainability of the IRS resulting in a less number of IRS elements for improving the system sum-rate via signal reflection.

\begin{figure}[t] \vspace*{-5mm}
  \centering
  \includegraphics[width=3.5in]{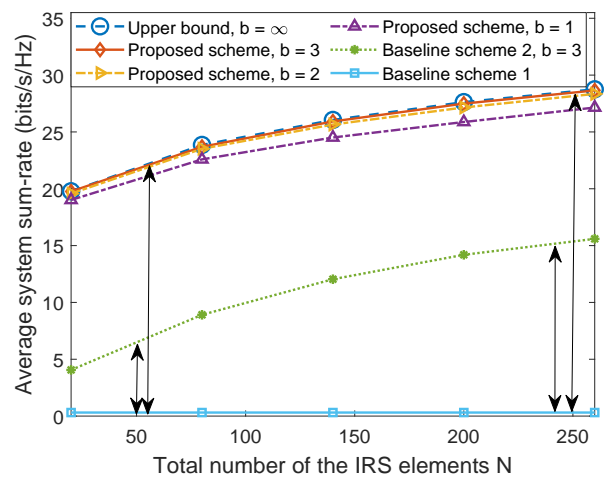}  \vspace*{-3mm}
  \caption{Average system sum-rate (bits/s/Hz) versus the total number of IRS elements.}
  \label{results:numN} \vspace*{-4mm}
\end{figure}

Fig. \ref{results:numN} shows the variation of average system sum-rate with different numbers of IRS elements $N$ at $d = 15$ m. It can be observed that with an increasing number of IRS elements $N$, the average system sum-rate of the proposed scheme increases. In particular,  a significant sum-rate gain can be achieved by the proposed scheme compared to baseline scheme $1$, even though the self-sustainability of the IRS is taken into account. Indeed, the extra spatial degrees of freedom offered by the increased number of reflecting IRS elements provides a higher flexibility in beamforming to enhance the channel quality of the end-to-end AP-IRS-user link for improving the system sum-rate.  Moreover, Fig. \ref{results:numN} also compares the performance of the proposed scheme with different bit resolutions, $b$, of IRS phase shifters. Note that the upper bound in Fig. \ref{results:numN} is the previously mentioned upper bound scheme but with $b=\infty$. It can be observed that the performance of the IRS phase shifts with a bit resolution at $b = 2,3$ bits approaches the upper bound. In particular, increasing bit resolution above $2$ bits would only provide a marginal improvement of system sum-rate. In fact, the IRS-user links are dominated by  LoS components in Rician fading channels. Therefore, a small bit resolution of phase shifts is sufficient to facilitate the beamformer aligning the desired signals with the dominant channels. Therefore, the bit resolution can be set as small as $2$ bits for practical low complexity designs.\vspace*{-3mm}
\section{Conclusions}
In this paper, we proposed the application of a self-sustainable IRS to a multi-user MISO downlink communication system. The joint design of the beamformer at the AP and the phase shifts and the power harvesting schedule at the IRS was formulated as a non-convex optimization problem to maximize the system sum-rate. Alternating optimization, SCA, and SDR techniques were employed to obtain a suboptimal solution of the design problem. Simulation results demonstrated that the proposed scheme offers significant performance gain compared to the conventional MISO system without IRS. Moreover, our results also unveiled the non-trivial trade-off between achieving self-sustainability and the system sum-rate. Lastly, we confirmed that a small number of bit resolution of phase shifters at IRS can achieve a considerable average system sum-rate of the ideal case with continuous phase shifters.\vspace{-1mm}

\bibliographystyle{IEEEtran}
\bibliography{IRS-EH}

\end{document}